%% file: main.tex
\newif\ifanonymousversion
\anonymousversionfalse

\documentclass[sigconf]{acmart}

\setcopyright{acmcopyright}
\renewcommand\footnotetextcopyrightpermission[1]{} 

\usepackage{graphicx}
\usepackage{xcolor}
\usepackage{soul}
\usepackage{caption}
\usepackage{subcaption}
\usepackage{multirow}
\usepackage{multicol}
\usepackage{adjustbox}
\usepackage{hyperref}

\newcommand{\nsf}[1]{\href{https://www.nsf.gov/awardsearch/showAward?AWD_ID=#1}{#1}}

\begin{document}

\title[Time Traveling to Defend Adversarial Example Attacks in Image Classification]{Time Traveling to Defend Against\\ Adversarial Example Attacks in Image Classification\vspace{1cm}}

\ifanonymousversion
\author{Anonymous Submission}

\else

\author{Anthony Etim}
\affiliation{%
  \institution{Yale University}
  \city{New Haven} 
  \state{CT} 
  \country{USA}
}
\email{anthony.etim@yale.edu}

\author{Jakub Szefer}
\affiliation{%
  \institution{Northwestern University}
  \city{Evanston} 
  \state{IL} 
  \country{USA}
}
\email{jakub.szefer@northwestern.edu}

\fi

\begin{abstract}
    \input{sections/abstract.tex}
\end{abstract}

\maketitle

\pagestyle{plain}


\input{sections/introduction.tex}

\input{sections/background}

\input{sections/threat_model.tex}

\input{sections/attack_implementation}

\input{sections/experimental_results}

\input{sections/discussion}

\input{sections/related_work}

\input{sections/conclusion.tex}

\input{sections/acknowledgements}

\bibliographystyle{ACM-Reference-Format}
\bibliography{bibtex/references}

\end{document}

%% file: sections/abstract.tex
Adversarial example attacks have emerged as a critical threat to machine learning. Adversarial attacks in image classification abuse various, minor modifications to the image that confuse the image classification neural network -- while the image still remains recognizable to humans. One important domain where the attacks have been applied is in the automotive setting with traffic sign classification. Researchers have demonstrated that adding stickers, shining light, or adding shadows are all different means to make machine learning inference algorithms mis-classify the traffic signs. This can cause potentially dangerous situations as a stop sign is recognized as a speed limit sign causing vehicles to ignore it and potentially leading to accidents. To address these attacks, this work focuses on enhancing defenses against such adversarial attacks. This work shifts the advantage to the user by introducing the idea of leveraging historical images and majority voting. While the attacker modifies a traffic sign that is currently being processed by the victim's machine learning inference, the victim can gain advantage by examining past images of the same traffic sign. This work introduces the notion of ``time traveling'' and uses historical Street View images accessible to anybody to perform inference on different, past versions of the same traffic sign. In the evaluation, the proposed defense has 100\% effectiveness against latest adversarial example attack on traffic sign classification algorithm.

%% file: sections/introduction.tex
\section{Introduction}
\label{sec_introduction}

Machine learning algorithms such as Deep Neural Networks (DNNs) are inherently vulnerable to adversarial example attacks, where small input modifications -- changes often imperceptible to the human eye -- can cause a trained model to incorrectly classify the input images~\cite{chakraborty2018adversarial}. Szegedy et al. first discovered that well-performing deep neural networks are susceptible to adversarial attacks~\cite{szegedy2013intriguing}. As these models are increasingly deployed in critical systems such as automated vehicles and surveillance, the potential for adversarial manipulation presents a serious threat to their reliability and safety especially in traffic sign classification and detection.

Successful adversarial attacks have been demonstrated on both traffic sign classification and detection models~\cite{eykholt2018robust, wei2022adversarial}, taking various forms. An attacker can manipulate the entire surface of a traffic sign~\cite{lu2017adversarial}, create adversarial stickers~\cite{eykholt2018robust}, use shadows~\cite{zhong2022shadows} or use light~\cite{hsiao2024natural} to disrupt the model's performance. These techniques exploit vulnerabilities in vision models, allowing subtle modifications that significantly impact traffic sign classification.

Despite growing concerns over adversarial attacks on traffic sign classification, few effective defense strategies have been developed. Common defenses like adversarial training~\cite{tramer2017ensemble, wan2020effects} and input preprocessing~\cite{qiu2020review} attempt to improve robustness by either exposing the model to adversarial examples during training or filtering adversarial noise, but these methods often fail to generalize across diverse attack types and real-world conditions. The limitations of existing defenses motivate this work's exploration of new ways of thinking about the attacks, and what information the victim can leverage to shift the balance of power in defending the attacks.

Our proposed defense method takes a novel approach by leveraging previous, historical images of a traffic sign and comparing them with the current input image of the traffic sign. The past images of the traffic sign can be analyzed, alongside the current input. The machine learning inference can be performed on the past images, and the current image, and majority voting can be performed to decide if the inference results are correct or not. Our method can effectively detect subtle adversarial manipulations, improving model resilience in both visible and imperceptible attack scenarios. Our data-driven approach enhances the model's ability to recognize traffic signs correctly and mitigates adversarial threats in real-time, providing a practical defense for traffic sign classification systems.

The defense is base on idea of ``time traveling'' to examine past images of traffic signs. The time traveling approach is enabled by widely available Street View images of traffic signs. Street View is a feature in Google Maps and Google Earth that offers interactive panoramic views from streets around the world, and the images clearly capture street signs. Street View can be accessed manually, or via an API. Especially, given a set of coordinates, the API can automatically retrieve the images from that location. In the case of traffic sign classification, for example, a self driving car can use its current coordinates to access the API and obtain images of the traffic sign it is currently facing from Street View. Importantly, Street View contains images of streets, and street signs, dating back to 2007 when the service was started. In addition to Google Street View, there are also services like Apple Look Around, Microsoft Bing Streetside, and a number of platforms with open-source software and crowd-sourced images.

\subsection{Contributions}

The contributions of this work are as follows:

\begin{enumerate}

    \item We develop a modified adversarial example attack to demonstrate the adversarial attack on real-life traffic signs from Street View images as opposed to attacking traffic sign images from a data set.
    
    \item To counter the attacks, we propose a novel defense method that compares historical images of traffic signs with current inputs and uses majority voting to detect adversarial manipulations in real-time.
    
    \item We integrate the defense with a traffic sign classification software to demonstrate the effectiveness of the defense.
    
    \item We provide extensive evaluations using numerous traffic sign images and years of past data from Street View.
    
    \item We discuss limitations of the approach and potential further defense mechanisms.
    
\end{enumerate}

%% file: sections/background.tex
\section{Background}
\label{background}

In this section, we provide brief overview of adversarial attacks as well as the LISA dataset which contains images of street signs in U.S. We also discuss LISA-CNN which is used in this work as the example of a victim machine learning model for street sign classification. In our work, we apply LISA-CNN, but use it with current Street View images of traffic signs as well as the historical Street View images of traffic signs.

\subsection{Adversarial Attacks}

Machine learning and DNN predictions can be altered through small pixel modifications in the input, a tactic known as adversarial attacks. These attacks can be executed in two primary settings: white-box, where the attacker has full access to the model’s architecture, weights, and data, or black-box, where the attacker has no knowledge of the target model~\cite{chakraborty2018adversarial,liang2022adversarial}.

One of the earliest and simplest white-box attack techniques is the Fast Gradient Sign Method (FGSM)~\cite{goodfellow2014explaining}. FGSM generates adversarial perturbations by adding the sign of the gradient of the loss function, scaled by a small constant, to the input image. This method has since been extended into more advanced iterative approaches, such as the Projected Gradient Descent (PGD) attack~\cite{mkadry2017towards}.

Adversarial attacks are typically performed on a per-instance basis, where each input image is perturbed with a unique noise pattern. However, a more practical and scalable method is the method in which a single perturbation pattern is crafted to attack all images within a dataset~\cite{moosavi2017universal}. This universal approach enhances the realism and applicability of adversarial attacks in real-world scenarios.

Further advancing the practicality of these attacks is the concept of adversarial patches, which localize the perturbations to a specific region of the image rather than modifying the entire image~\cite{brown2017adversarial}. These patches are designed in a universal manner, making them suitable for real-world applications. 

To improve the robustness of adversarial patches against spatial shifts, varying camera angles, and environmental changes, several techniques have been developed. In ~\cite{athalye2018synthesizing}, the authors introduced the Expectation over Transformations (EoT) framework, which applies a variety of transformations—such as rotation, scaling, and brightness adjustments—during training to enhance patch resilience. However, Eykholt et al.~\cite{eykholt2018robust} noted that the synthetic transformations in ~\cite{athalye2018synthesizing} fail to fully capture the complexities of real-world conditions. Instead, they proposed the Robust Physical Perturbations (RP2) method, which incorporates real-world variability, such as images taken from different angles, distances, and lighting conditions, providing a more practical solution.

In addition to these methods, light and shadow-based attacks have emerged as new adversarial techniques. 
In~\cite{zhong2022shadows}, the  authors leveraged shadows to create disruptive perturbations on target objects. They showed that that this attack was effective in both digital and physical environments and posed a serious threat by causing machine learning-based vision systems to make incorrect decisions. Unlike traditional noise-based attacks, shadow attacks are particularly stealthy, as they harness a common natural phenomenon that can easily bypass both human observers and AI-based defenses. This makes them a formidable challenge for real-world AI systems.

In~\cite{hsiao2024natural}, the authors  demonstrated that everyday natural light sources, such as sunlight and flashlights, can pose a significant threat to image classification tasks. This attack can be  carried out by anyone, exposing the vulnerability of vision systems to common environmental lighting conditions. These attacks exploit natural phenomena like shadows or artificial light to create subtle perturbations that are difficult for both humans and machine learning models to detect. By manipulating lighting conditions, adversaries can craft imperceptible changes that mislead models, adding an extra layer of complexity to defending against real-world adversarial~threats.

Our defense can be applied to protect against these attacks. It leverages the fact that the adversary most often has access only to the current traffic sign being classified. For example, the adversary modifies the traffic sign that is being classified, e.g., by adding shadows to it. Our defense leverages the fact that the victim can shift the balance of power by using historical images of the same traffic sign in a separate image processing pipeline to classify the past images, and compare the results to the classification of the current traffic sign.

\subsection{LISA-CNN and Traffic Sign Dataset}

LISA is a U.S. traffic sign dataset that comprises of 47 different road signs~\cite{lisa}. Following previous research, our work focuses on the 16 most common signs to address the dataset's unbalanced distribution~\cite{hsiao2024natural}. This selection allows for more effective training of convolutional neural networks (CNNs) tailored for traffic sign classification tasks. The LISA-CNN architecture consists of three convolutional layers followed by a fully connected layer, specifically trained to classify images from the LISA dataset~\cite{eykholt2018robust}. This architecture leverages the diverse range of images captured under various environmental conditions and perspectives, making it essential for developing robust models capable of generalizing across real-world scenarios. Studies have shown that models trained on the LISA dataset exhibit high accuracy and real-time performance in detecting and classifying traffic signs, thereby advancing the capabilities of autonomous vehicle perception systems~\cite{pavlitska2023adversarial}.

Our defense leverages the LISA-CNN since it is pre-trained on U.S. traffic signs. LISA-CNN can be used to process current (possibly malicious) input, as well as historical street sign images.

\begin{figure*}[t]
    \begin{subfigure}[b]{0.19\textwidth}
        \centering
\includegraphics[width=2.2cm]{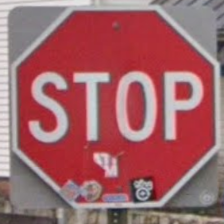}
        \caption{\small \centering Stop\newline Test Image}
        \label{fig:stop_image}
    \end{subfigure}
    \hfill
     \begin{subfigure}[b]{0.19\textwidth}
        \centering
    \includegraphics[width=2.2cm]{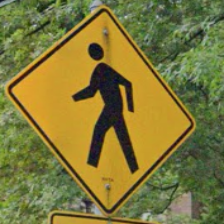}
        \caption{\small \centering Ped. Crossing\newline Test Image}
    \label{fig:pedestrian_image}
    \end{subfigure}
    \hfill
    \begin{subfigure}[b]{0.19\textwidth}
        \centering
        \includegraphics[width=2.2cm]{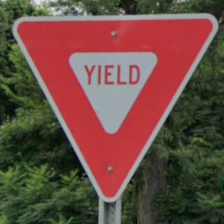}
        \caption{\small \centering Yield\newline Test Image}
    \label{fig:yield_image}
    \end{subfigure}
    \hfill
    \begin{subfigure}[b]{0.19\textwidth}
        \centering
        \includegraphics[width=2.2cm]{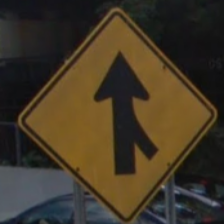}
        \caption{\small \centering Merge\newline Test Image}
    \label{fig:merge_image}
    \end{subfigure}
    \hfill
    \begin{subfigure}[b]{0.19\textwidth}
        \centering
        \includegraphics[width=2.2cm]{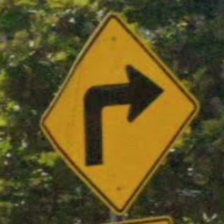}
        \caption{\small \centering Turn Right\newline Test Image}
    \label{fig:right_image}
    \end{subfigure}
    \hfill
    \caption{Test images used in evaluation of the attacks and defenses.}
    \label{fig:Test_Images}
\end{figure*}

\begin{figure*}[t]
    \begin{subfigure}[b]{0.19\textwidth}
        \centering
\includegraphics[width=2.2cm]{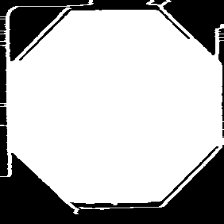}
        \caption{\small \centering Stop\newline Output Mask}
        \label{fig:stop_output_mask}
    \end{subfigure}
    \hfill
     \begin{subfigure}[b]{0.19\textwidth}
        \centering
    \includegraphics[width=2.2cm]{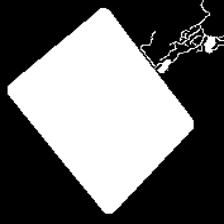}
        \caption{\small \centering Ped. Crossing\newline Output Mask}
    \label{fig:pedestrian_output_mask}
    \end{subfigure}
    \hfill
    \begin{subfigure}[b]{0.19\textwidth}
        \centering
        \includegraphics[width=2.2cm]{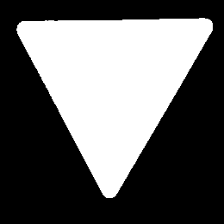}
        \caption{\small \centering Yield\newline Output Mask}
    \label{fig:yield_output_mask}
    \end{subfigure}
    \hfill
    \begin{subfigure}[b]{0.19\textwidth}
        \centering
        \includegraphics[width=2.2cm]{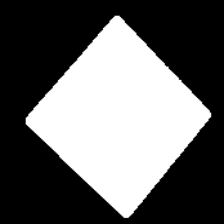}
        \caption{\small \centering Merge\newline Output Mask}
    \label{fig:merge_output_mask}
    \end{subfigure}
    \hfill
    \begin{subfigure}[b]{0.19\textwidth}
        \centering
        \includegraphics[width=2.2cm]{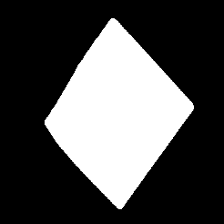}
        \caption{\small \centering Turn Right\newline Output Mask}
    \label{fig:right_output_mask}
    \end{subfigure}
    \hfill
    \caption{Shadow masks generated by our enhanced attack method, one for each test image.}
     \label{fig:Output Masks}
\end{figure*}

%% file: sections/threat_model.tex
\section{Threat Model}

In our threat model, we assume that the attacker has partial knowledge of the machine learning system but does not have direct access to the model’s internal architecture, weights, or training data, nor digital images stored by the victim, fitting a black-box attack scenario. The adversary’s goal is to disrupt the system's performance by exploiting vulnerabilities in the physical environment, such as manipulating natural light sources or shadows, without requiring physical contact with the target object or system. The attacker can control external factors like light direction, intensity, and angle to introduce subtle perturbations that are imperceptible to human observers but sufficient to mislead the model into making incorrect predictions. We further assume that the adversary is capable of deploying these attacks in real-world settings, such as autonomous vehicles or surveillance systems, where image classification plays a critical role in decision-making. The attack is designed to bypass traditional defenses against adversarial noise, relying instead on environmental manipulation to achieve the desired effect. We assume the attacker is not able to manipulate other information, such as GPS signal, that may confuse the autonomous vehicle about its location. We assume the victim, e.g., the autonomous vehicle, can use the internet to access Street View images based on its location, or has pre-loaded historical maps and Street View images if internet access is not possible.

%% file: sections/attack_implementation.tex
\section{Adversarial Attack on Street Signs}

This section presents an adversarial attack on traffic sign classification. Existing shadow attack is extended to work with Street View and the new extended attack is evaluated on real-life Street View~images.

\subsection{Shadow Attack}
\label{shadow_attack}

In this work, we start with previously introduced shadow-based adversarial attack, originally proposed by Zhong et al.~\cite{zhong2022shadows}, which utilizes shadows as a natural and stealthy form of perturbation. Our focus, however, is to apply the attack to Street View images as an example of an attack on real-life traffic signs.

The shadow attack, leveraging natural phenomena of light shadows, can subtly manipulate machine learning models in ways that are difficult to detect. The shadow attack leverages Particle
Swarm Optimization (PSO)~\cite{kennedy1995particle} approach to find most optimal location to add a shadow to an image. A shadow is a polygon of dark or gray color that is meant to look like a shadow cast by a tree, building, or another object. The attack analyzes different shadow shapes and locations in order to determine the optimal type of shadow. Once the best shadow is determined, the attacker can reproduce it by applying the same shadow to the real street sign (if manipulating a physical street sign that is later captured and processed by the LISA-CNN street sign categorization algorithm) or to a digital image of the street sign (if manipulating a digital representation of the street sign before it is processed by LISA-CNN). The limitation of the existing attack is that for each image, there is a need to define a mask, or region where the shadow is applied. For real-life images, the mask has to be generated on the fly for each image as even if the shape of the street sign is known, the angle from which it is imaged will affect the mask.

\begin{figure*}[t]
    \begin{subfigure}[b]{0.19\textwidth}
        \centering
\includegraphics[width=2.2cm]{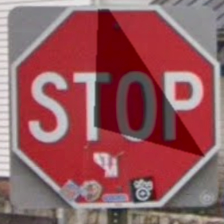}
        \caption{\small \centering Stop Adversarial\newline Image}
        \label{fig:stop_adversarial_image}
    \end{subfigure}
    \hfill
     \begin{subfigure}[b]{0.19\textwidth}
        \centering
    \includegraphics[width=2.2cm]{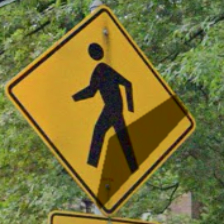}
        \caption{\small \centering Ped. Crossing Adversarial Image}
    \label{fig:pedestrian_adversarial_image}
    \end{subfigure}
    \hfill
    \begin{subfigure}[b]{0.19\textwidth}
        \centering
        \includegraphics[width=2.2cm]{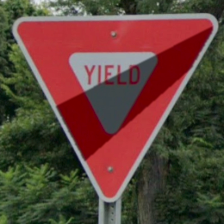}
        \caption{\small \centering Yield Adversarial\newline Image}
    \label{fig:yield_adversarial_image}
    \end{subfigure}
    \hfill
    \begin{subfigure}[b]{0.19\textwidth}
        \centering
        \includegraphics[width=2.2cm]{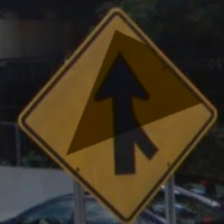}
        \caption{\small \centering Merge Adversarial\newline Image}
    \label{fig:merge_adversarial_image}
    \end{subfigure}
    \hfill
    \begin{subfigure}[b]{0.19\textwidth}
        \centering
        \includegraphics[width=2.2cm]{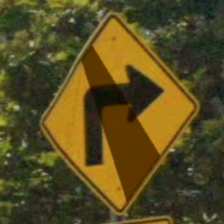}
        \caption{\small \centering Turn Right Adversarial\newline Image}
    \label{fig:right_adversarial_image}
    \end{subfigure}
    \hfill
    \caption{Attack images which include the adversarial shadows added within the previously computed mask regions.}
    \label{fig:Adversarial Images}
\end{figure*}

\subsection{Custom Shadow Masks}

In our modified approach, we developed a custom shadow attack that required the generation of binary masks for each input traffic image so that realistic shadows can be applied. These masks were crucial in defining the areas where adversarial perturbations would occur, creating more naturalistic and effective attacks. To automate the mask generation, we processed each traffic image by detecting the most prominent contours and using them to create a refined binary mask.

The mask generation process begins by reading the input image and converting it to grayscale for easier processing. A Gaussian blur~\cite{hummel1987deblurring} is applied to reduce noise, followed by Canny edge detection algorithm~\cite{rong2014improved} to identify the edges within the image. After edge detection, dilation and morphological closing techniques~\cite{gil2002efficient} are used to fill gaps in the contours, ensuring a more cohesive mask. Valid contours are then filtered based on their area, and the largest contour is selected to generate the mask. Finally, this contour is drawn onto a black background, creating a binary mask that highlights the target area where the shadow perturbations will be applied to each sign.

This method enabled us to customize masks for each traffic sign image, regardless of the angle from which it is captured and ensuring that the shadow attack was applied in a precise and controlled manner. The flexibility of this approach allows for accurate generation of real-world shadow effects, further enhancing the stealthiness of the adversarial examples.

Demonstration of our masks generation is shown in Figures~\ref{fig:Test_Images}  and~\ref{fig:Output Masks}. We start with 5 sample current or most-recent street sign images captured from Street View. The images consider different angles of viewing the signs and different lighting conditions. The signs are from U.S. but specific location is not disclosed to maintain anonymity of the authors. The 5 test images are shown in Figure~\ref{fig:Test_Images}. To implement the enhanced attack, we generate the masks, as shown in Figure~\ref{fig:Output Masks}.

\subsection{Attack on Street View Images}

We apply the adversarial attack to Street View images, specifically targeting traffic signs in the United States. The Street View images were chosen to reflect common conditions in U.S. environments since the model was trained on the LISA dataset, which contains U.S. traffic signs. This allowed us to explore how vulnerabilities in machine learning models, trained on localized datasets, can be exploited in broader real-world applications such as autonomous vehicle navigation and mapping services.

Google Street View images~\cite{anguelov2010google}, captured at various angles, lighting conditions, and environmental settings, provide a challenging real-world scenario for both attack and defense strategies. These images, which are critical for navigation and safety in autonomous systems, present unique opportunities for adversarial attacks to exploit vulnerabilities in object classification algorithms.

For this investigation, we applied our enhanced shadow-based adversarial attack method mention in Section~\ref{shadow_attack} to manipulate the classification of traffic signs in Street View images. This attack leverages natural shadows as perturbations, making it particularly difficult to detect, both by humans and automated systems. Unlike other perturbation-based attacks like pixel-level noise or adversarial patches, shadows blend seamlessly with real-world environments, allowing for highly stealthy and effective adversarial examples.

\input{sections/tables/table_adversarial}

The adversarial images with the shadows can be seen in Figure~\ref{fig:Adversarial Images}. For each image, a shadow is generated that causes the image to be mis-classified. Table~\ref{table_adversarial} show the results of the attack. For each image, after shadow is added, it is classified incorrectly with high confidence score. Thus, real-life street signs can be modified with shadows and the result will confuse autonomous vehicles, for example, to merge instead of yield at a pedestrian crossing.

%% file: sections/tables/table_adversarial.tex
\begin{table}[t]
\centering
\caption{Predicted labels of the adversarial images and their confidence scores.}
\label{table_adversarial}
\small
\begin{tabular}{|p{1.8cm}|p{1.8cm}|p{1.8cm}|p{1.2cm}|}
\hline 
\textbf{Adversarial Image} & \textbf{Predicted Label}  & \textbf{Confidence Score (\%)} & \textbf{Attack Success}
\\ \hline \hline
Stop &  Ped. Crossing &  84.67 & Yes\\ \hline 
Ped. Crossing  & Merge & 90.67 & Yes\\ \hline 
Yield  & Turn Right & 77.57 & Yes\\ \hline 
Merge & Ped. Crossing & 93.51 & Yes\\ \hline 
Turn Right & Ped. Crossing & 82.69 & Yes\\ \hline 
\end{tabular}
\end{table}

%% file: sections/experimental_results.tex
\input{sections/attack_results}

\input{sections/defense_results}

%% file: sections/attack_results.tex
\begin{figure*}[t]
    \begin{subfigure}[b]{0.33\textwidth}
        \centering
\includegraphics[width=2.2cm]{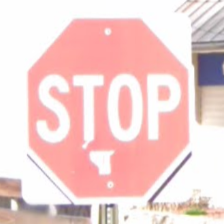}
        \caption{\small November 2020 Stop Image}
        \label{fig:stop_defense_image_1}
    \end{subfigure}
    \hfill
     \begin{subfigure}[b]{0.33\textwidth}
        \centering
\includegraphics[width=2.2cm]{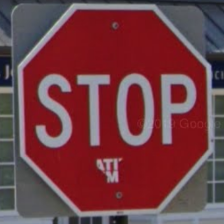}
        \caption{\small July 2019 Stop Image}
        \label{fig:stop_defense_image_2}
    \end{subfigure}
    \hfill
    \begin{subfigure}[b]{0.33\textwidth}
        \centering
\includegraphics[width=2.2cm]{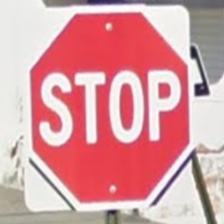}
        \caption{\small October 2016 Stop Image}
        \label{fig:stop_defense_image_3}
    \end{subfigure}
    \hfill
    \begin{subfigure}[b]{0.33\textwidth}
        \centering
\includegraphics[width=2.2cm]{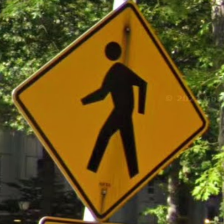}
        \caption{\small September 2022 Ped. Crossing Image}
\label{fig:pedestrian_defense_image_1}
    \end{subfigure}
    \hfill
     \begin{subfigure}[b]{0.33\textwidth}
        \centering
\includegraphics[width=2.2cm]{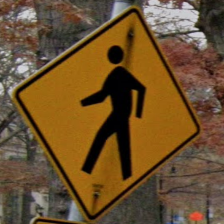}
        \caption{\small December 2021 Ped. Crossing Image}
\label{fig:pedestrian_defense_image_2}
    \end{subfigure}
    \hfill
    \begin{subfigure}[b]{0.33\textwidth}
        \centering
\includegraphics[width=2.2cm]{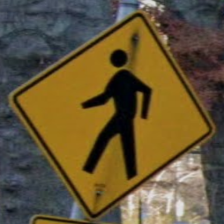}
        \caption{\small November 2020 Ped. Crossing Image}
\label{fig:pedestrian_defense_image_3}
    \end{subfigure}
    \hfill
    \begin{subfigure}[b]{0.33\textwidth}
        \centering
\includegraphics[width=2.2cm]{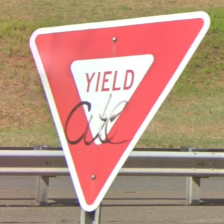}
        \caption{\small August 2022 Yield Image}
\label{fig:yield_defense_image_1}
    \end{subfigure}
    \hfill
     \begin{subfigure}[b]{0.33\textwidth}
        \centering
\includegraphics[width=2.2cm]{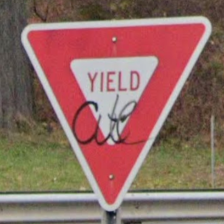}
        \caption{\small December 2021 Yield Image}
\label{fig:yield_defense_image_2}
    \end{subfigure}
    \hfill
    \begin{subfigure}[b]{0.33\textwidth}
        \centering
\includegraphics[width=2.2cm]{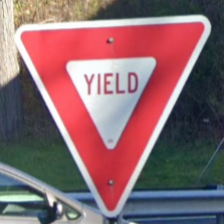}
        \caption{\small November 2020 Yield Image}
\label{fig:yield_defense_image_3}
    \end{subfigure}
    \hfill
    \begin{subfigure}[b]{0.33\textwidth}
        \centering
\includegraphics[width=2.2cm]{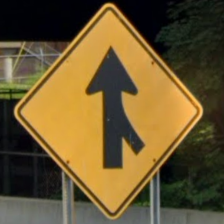}
        \caption{\small July 2019 Merge Image}
\label{fig:merge_defense_image_1}
    \end{subfigure}
    \hfill
     \begin{subfigure}[b]{0.33\textwidth}
        \centering
\includegraphics[width=2.2cm]{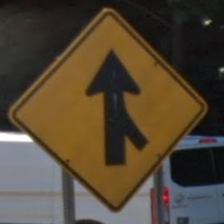}
          \caption{\small July 2018 Merge Image}
\label{fig:merge_defense_image_2}
    \end{subfigure}
    \hfill
    \begin{subfigure}[b]{0.33\textwidth}
        \centering
\includegraphics[width=2.2cm]{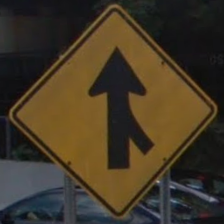}
          \caption{\small September 2017 Merge Image}
\label{fig:merge_defense_image_3}
    \end{subfigure}
    \hfill

    \begin{subfigure}[b]{0.33\textwidth}
        \centering
\includegraphics[width=2.2cm]{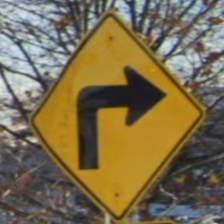}
        \caption{\small December 2021 Right Image}
\label{fig:right_defense_image_1}
    \end{subfigure}
    \hfill
     \begin{subfigure}[b]{0.33\textwidth}
        \centering
\includegraphics[width=2.2cm]{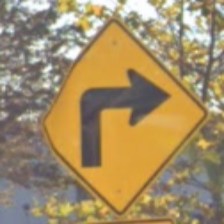}
        \caption{\small November 2020 Right Image}
\label{fig:right_defense_image_2}
    \end{subfigure}
    \hfill
    \begin{subfigure}[b]{0.33\textwidth}
        \centering
\includegraphics[width=2.2cm]{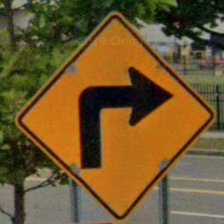}
        \caption{\small July 2019 Right Image}
\label{fig:right_defense_image_3}
    \end{subfigure}
    \hfill
    \caption{Historical images of the 5 test street signs used in evaluating the defense. Historical data is not always available for the same dates for each image, thus 3 prior images of each sign were used, resulting in some images being from different dates.}
      \label{fig:Defense Images}
\end{figure*}

%% file: sections/defense_results.tex
\section{Defense: Time Traveling to Defend Against Adversarial Example Attacks}

To enhance the robustness of traffic sign classification models against adversarial attacks, we introduce a novel defense mechanism utilizing historical image data from Street View. Our ``time travel'' approach involves leveraging past historical images of traffic signs, and comparing current (possibly) adversarial traffic sign images to their past versions. By leveraging well-known majority voting approach, image classification of past traffic sign images and the current image can be compared to determine if the image may have been manipulated to cause incorrect classification.

\subsection{Accessing Past Traffic Sign Images}

Both manual and programmatic methods can be used to access Street View images. Given a location or coordinates, Street View allows for viewing current and past images. Manually, the Street View is available through a web page, and screenshots of the street sign images can be captured. Generally, web page Street View provides higher resolution images. For programmatic approach, we utilized a Python package~\cite{street_view} for retrieving current and historical photos from Google Street View, which provides a convenient means to access Street View imagery and build our dataset. However, the API often yields low-resolution outputs that may not capture the intricate details necessary for accurate analysis and classification. Thus depending on the street sign, either manual or programmatic method needs to be used to obtain the best image for use with the image classification algorithms.

\input{sections/tables/table_attack_summary}

\input{sections/tables/table_historical}

A side-benefit of using Street View is that images of street signs are generally taken on clear, sunny days, thus they can be classified with high accuracy.

\subsection{Classification of Past Traffic Sign Images}

We first analyze our ``time travel'' approach to classify prior images of each of the 5 test street signs. Using the location or coordinates of the street sign, we retrieve the past images from Street View. The images are shown in Figure~\ref{fig:Defense Images}. Historical data is not always available for the same dates for each image, thus 3 prior images of each sign were retrieved. This actually allows for testing images from many years past, in our case dating back to 2016. In general, Street View is available from 2007 in U.S. thus much historical data is available for many street signs.

Table~\ref{table_historical} shows the LISA-CNN classification of the past images. In the table, we present the model’s performance when classifying historical traffic sign images under no defense, along with the corresponding confidence scores. The model consistently predicts the correct label across all traffic sign categories, indicating that past images can be used to obtain correct classification when no attack is present. Turn right signs show the best classification, with confidence scores reaching as high as 94.85\%. Stop signs yield confidence scores between 77.02\% and 87.40\%, suggesting that even for older images, such as from 2016, the model remains confident in its predictions despite adversarial manipulation. Similarly, pedestrian crossing signs are accurately classified with confidence scores ranging from 76.56\% to 87.17\%. However, yield signs exhibit a slight drop in confidence, with scores as low as 68.30\%.

\subsection{Defense Results}

Having determined that past traffic sign images are consistently and correctly classified, we leverage them in majority voting scheme to defend against the adversarial attacks. For each image, we consider the current image (under attack) and 3 past images obtained based on location or coordinates of the current traffic sign. Classification is performed on 4 images (current image plus 3 past images) and majority voting is used to determine the predicted label.

The results are shown in Table~\ref{table_attack_summary}. We can see that with 100\% success rate, the defense prevents adversarial attacks. Ubiquitous access to Street View in U.S. and other countries means that our defense can be widely deployed. Considering autonomous vehicles, if the route is known ahead of time, the past images can be retrieved before the vehicle reaches the next street sign, thus classification can be done ahead of time, and not add overhead when current street sign is being classified, except for the majority voting computation that is trivial.

In order to compare to prior defenses, we have tested the adversarial images with the LISA-CNN model trained on the adversarial shadow images~\cite{zhong2022shadows}. This method of adversarial training by adding the shadows in each training sample helps to decrease the success of the attack but it does not successfully defend against it in all cases. From Table~\ref{table_attack_summary}, we can see that this adversarial training defense does not defend against all the adversarial images. This highlights that our proposed  ``time traveling'' defense is more robust, as it leverages historical images to detect inconsistencies in real-time scene analysis, offering improved resilience against adversarial attacks.

%% file: sections/tables/table_attack_summary.tex
\begin{table*}[t]
\centering
\caption{Comparison of predicted labels with and without defenses.}
\label{table_attack_summary}
\small
\begin{tabular}{|p{2.3cm}|p{2.8cm}|p{1.3cm}|p{2.8cm}|p{1.3cm}|p{2.8cm}|p{1.3cm}|}
\hline
                    & \multicolumn{2}{c|}{\textbf{No Defense}}  
                    & \multicolumn{2}{c|}{\textbf{Defense~\cite{zhong2022shadows}}}  
                    & \multicolumn{2}{c|}{\textbf{Our Defense}}           \\ \hline
\textbf{Image}  & \textbf{Predicted Label}  & \textbf{Defense Succeeded}  & \textbf{Predicted Label}  & \textbf{Defense Succeeded} & \textbf{Predicted Label}  & \textbf{Defense Succeeded}  \\ \hline
Stop                & Ped. Crossing & $\times$   & Stop & \checkmark                 & Stop & \checkmark                       \\ \hline
Ped. Crossing & Merge               & $\times$    &Merge & $\times$                 & Ped. Crossing              & \checkmark                      \\ \hline
Yield               & Turn Right          & $\times$   & Yield & \checkmark                 & Yield         & \checkmark                       \\ \hline
Merge               & Ped. Crossing & $\times$    &Ped. Crossing & $\times$                  & Merge & \checkmark                      \\ \hline
Turn Right              & Ped. Crossing & $\times$  &Ped. Crossing & $\times$ & Turn Right &  \checkmark \\ \hline
\end{tabular}
\end{table*}

%% file: sections/tables/table_historical.tex
\begin{table}[t]
\centering
\caption{Predicted labels of the historical images and their confidence scores.}
\begin{adjustbox}{width=0.48\textwidth}
\label{table_historical}
\small
\begin{tabular}{|p{2.2cm}|p{1.5cm}|p{2cm}|p{1.6cm}|}
\hline 
\textbf{Image} & \textbf{Date} &\textbf{Predicted Label}  & \textbf{Confidence Score (\%)}
\\ \hline \hline
Stop & Nov. 2020 &  Stop &  77.02\\ \hline 
 Stop     &Jul. 2019 & Stop & 85.76\\ \hline 
Stop  &Oct. 2016  & Stop & 87.40\\ \hline 
\hline
 Ped. Crossing & Sep. 2022& Ped. Crossing & 87.17\\ \hline
 Ped. Crossing  & Dec. 2021& Ped. Crossing & 76.56\\ \hline
 Ped. Crossing  & Nov. 2020& Ped. Crossing & 86.74\\ \hline
\hline
Yield    &Aug. 2022  & Yield & 86.86\\ \hline
 Yield  & Dec. 2021& Yield & 70.45\\ \hline
Yield  &Nov. 2020  & Yield & 68.30\\ \hline
\hline
Merge      & Jul. 2019 & Merge & 82.26\\ \hline
 Merge      &Jul. 2018 & Merge & 71.90\\ \hline
 Merge & Sep. 2017& Merge & 82.58\\ \hline
\hline
 Turn Right & Dec. 2021& Turn Right & 94.85\\ \hline
 Turn Right &Nov. 2020 & Turn Right & 87.50\\ \hline
Turn Right     &Jul. 2019  & Turn Right & 92.78\\ \hline

\end{tabular}
\end{adjustbox}
\end{table}

%% file: sections/discussion.tex
\section{Discussion and Limitations}

The proposed defense can be easily deployed. It does, however, depend on the ``time travel'' approach and accessing past images of street signs. If such data is not available, then the defense cannot be deployed, e.g., in countries where Street View is not available. When Street View is available, other problems are possible such as a newly-installed street sign that has no historical data. Defense can be in this augmented to generate a warning if not enough past images of a street sign are available. Similarly, a street sign can change, such as speed limit is increased or decreased on a road. Defense can be in this case augmented to issue a warning if current classification does not match past classification,  but this may not mean an attack, but that some street sign configuration changed.

%% file: sections/related_work.tex
\section{Related Work}
\label{sec:related_work}

This section provides an overview of existing work on defenses against adversarial example attacks. Distillation, as proposed by Hinton et al. \cite{hinton2015distilling}, is a method that facilitates the transfer of knowledge from complex neural networks to simpler architectures. To address adversarial attacks, Papernot et al. \cite{papernot2016distillation} introduced a defensive distillation strategy. This technique first trains a distillation model using the original inputs and labels to generate a probability distribution. This distribution, combined with the original examples, is then used to train a new model of the same architecture, yielding an updated probability distribution as the new label enhancing the model's robustness against adversarial threats. Liao et al.~\cite{liao2018defense} proposed a high-level representation guided denoiser (HGD) that utilizes a U-Net architecture as the denoising network. This approach incorporates a loss function targeting high-level features, suppressing error amplification and improving model performance. 

To prevent attackers from exploiting gradient information to attack the model, the authors in~\cite{folz2020adversarial} proposed a defense mechanism using S2SNet, which masks the model's gradient. S2SNet transforms category-related information into structural data that impacts the gradient, then encodes only the essential structural components for the classification task, discarding irrelevant parts to mitigate adversarial perturbations.

In ~\cite{wu2019defending}, the authors demonstrated that adversarial training with PGD attacks~\cite{madry2017towards} and randomized smoothing are limited in their effectiveness against high-profile physical attacks. They introduced a novel adversarial model, Rectangular Occlusion Attacks (ROA), coupled with adversarial training where an adversary inserts a small, adversarially crafted rectangle into an image. They showed that adversarial training with this new attack significantly enhances the robustness of image classification models against physically realizable attacks.

Cohen et al.~\cite{cohen2020detecting} introduced a method for detecting adversarial attacks on any pre-trained neural network. Using influence functions and a k-nearest neighbor (k-NN) model on the DNN's activation layers, they ranked supportive training samples. Normal inputs showed strong correlations with their nearest neighbors, while adversarial inputs did not successfully catch adversarial examples.

%% file: sections/conclusion.tex
\section{Conclusion}
\label{sec_conclusion}

This work highlights the vulnerability of traffic sign classification models to adversarial attacks, specifically through the implementation of shadow attacks using real-life Street View images. By analyzing the impact of these attacks, we demonstrated the necessity for a new defense mechanism to ensure the reliability of machine learning models in real-world applications. Our proposed approach, which utilizes historical Street View data and high-resolution images of street signs offers a promising avenue for enhancing resilience against adversarial perturbations attacks. By comparing adversarial traffic images with their historical counterparts, we leverage majority voting to prevent adversarial attacks. Ultimately, our findings underscore the importance of developing robust defense mechanisms against adversarial attacks.

%% file: sections/acknowledgements.tex
\section*{Acknowledgements}

This work was supported in part by National Science Foundation grant \nsf{2245344}.